\documentclass[12pt]{article} 
\usepackage{latexsym,amssymb,amsmath}
  \newtheorem{theorem}{Theorem}
  \newtheorem{proposition}{Proposition}
               
               \newtheorem{lemma}{Lemma}

               \def\pf{\par\noindent {\em Proof.}~\par\noindent}
               \def\qed{~\hfill{$\square$}\pagebreak[1]\par\medskip\par}

\newcommand{\R}{{\mathbb R}}

\newcommand{\I}{{\cal I}}
\newcommand{\uI}{{\underline{\cal I}}}

\newcommand{\dv}{{\rm div}}

\newcommand{\ux}{\underline{x}}
\newcommand{\pD}{{^\psi\!\upa}}
\newcommand{\hD}{{^\phi\!\upa}}

\newcommand{\og}{\overline{g}}
\newcommand{\upa}{\underline{\partial}}
\newcommand{\pux}{\partial_{\ux}}

\newcommand{\cH}{{\cal H}}
\newcommand{\ucH}{{{\underline{\cal H}}}}

\newcommand{\E}{{\bf E}}

\newcommand{\grad}{{\rm grad}}

\begin{document}
\title{On a Generalized Lam\'e-Navier system in $\R^3$}
\small{
\author
{Daniel Alfonso Santiesteban; Ricardo Abreu Blaya;\\Mart\'in Patricio \'Arciga Alejandre}
\date{Facultad de Matem\'aticas, Universidad Aut\'onoma de Guerrero, M\'exico.\\ Emails: danielalfonso950105@gmail.com, rabreublaya@yahoo.es, mparciga@gmail.com}

\maketitle
\begin{abstract}
This paper is devoted to a fundamental system of equations in Linear Elasticity Theory: the famous Lam\'e-Navier system. The Clifford algebra language allows us to rewrite this system in terms of the euclidean Dirac operator, which at the same time suggests a very natural generalization involving the so-called structural sets. We are interested in finding some structures in the solutions of these generalized Lam\'e-Navier systems. Using MATLAB we also implement algorithms to compute with such partial differential operators as well as to verify some theoretical results obtained in the paper.       
\end{abstract}

\vspace{0.3cm}

\small{
}
\section{Introduction}

In the state of equilibrium the three-dimensional displacement vector $\vec{u}$  should satisfy the Lam\'e-Navier system 
\begin{equation}\label{Lame}
\mu\triangle\vec{u}+(\mu+\lambda)\grad(\dv\vec{u})=0, 
\end{equation}
at any point within a homogeneous isotropic linear elastic body without volume forces. 
The quantities $\mu>0$ and $\lambda>-\frac{2}{3}\mu$ are called the Lam\'e constants.

This system was originally introduced by G. Lam\'e in 1837 \cite{La} while studying the method of separation of variables for solving the wave equation in elliptic coordinates. Moreover, its applications cover many branches in the fields such as linear elastostatics, chaotic Hamiltonian systems, and the theory of Bose-Einstein condensates \cite{Ba,Fung,Mal,Sadd,Sokol,JERR,Mu}. 

From \cite{MAB3} it is known that the Lam\'e equation (\ref{Lame}) can be rewritten in the form
\begin{equation}\label{lameoper}
\left(\frac{\mu+\lambda}{2}\right)\upa\vec{u}\upa+\left(\frac{3\mu+\lambda}{2}\right)\upa^2\vec{u}=0,
\end{equation}
where
\[
\upa:=e_1{\frac{\partial}{\partial x_1}}+e_2{\frac{\partial}{\partial x_2}}+e_3{\frac{\partial}{\partial x_3}}
\]
stands for the Dirac operator in $\R^3$ constructed with the generators $\{e_1,e_2,e_3\}$ of the real Clifford algebra $\R_{0,3}$. The null-solutions of $\pux$ are referred in the literature as monogenic functions \cite{BDS,GS}.
 
The search for all linear partial differential operators of the form

\begin{equation}\label{DiracPsi}
\pD:=\psi^1{\frac{\partial}{\partial x_1}}+\psi^2{\frac{\partial}{\partial x_2}}+\psi^3{\frac{\partial}{\partial x_3}},
\end{equation}
such that solutions of the differential equation $\pD u=0$ are always solutions of the Laplace equation $\Delta u=0$,  goes back to Nono \cite{No}. 

Let $\pD$ be a linear differential operator of the form \eqref{DiracPsi} with coefficients $\psi^i\in\R^3\subset\R_{0,3}$. To fulfill the Laplacian factorization $\pD\pD=-\Delta$ in $\R^3$, the following relations hold
\[\psi^i\psi^j+\psi^j\psi^i=-2\delta_{ij}\,\,(i,j=1,2,3).\]

The system $\{\psi^1,\psi^2,\psi^3\}$ can be thought of as an orthonormal (in the usual Euclidean sense) basis in $\R^3$. In this way, we obtain what will be referred to as structural set \cite{SV}.

The $\R_{0,3}$-valued solutions of $\pD u=0$ are the so-called $\psi$-hyperholomorphic functions. As pointed out in \cite{GN2}, the class of $\psi$-hyperholomorphic functions is wider than the one we get by rotations from the class of monogenic functions. The flexibility introduced by the structural sets allows us to look for new perspectives in several lines of research concerning the mapping properties of a related $\Pi$-operator, geometric conformal mappings and additive decompositions of harmonic functions \cite{A1,A2,DKM,G,GKSh,GN1,KM,N}.

It is precisely in this scenario that a generalization of the Lam\'e equation naturally emerges. Indeed, the idea is to consider in \eqref{lameoper} the generalized Dirac operator ${^\psi}\pux$ rather than the standard one. 

In line with that way of thinking, we arrive at two possible generalizations of the Lam\'e system \eqref{Lame}:
\begin{equation}\label{lamepsi}
\alpha[\pD\vec{u}\,\pD]+\beta[\pD\pD\vec{u}]=0
\end{equation}
and
\begin{equation}\label{lamephipsi}
\alpha[\hD\vec{u}\pD]+\beta[\hD\pD\vec{u}]=0,
\end{equation}
where $\phi,\psi$ are two structural sets and for brevity we used the notation $\alpha=\frac{\mu+\lambda}{2}$, $\beta=\frac{3\mu+\lambda}{2}$.

This paper aims to investigate the structure of the solutions of these generalized systems, as well as to determine the similarities and differences between them and the solutions of the classical Lam\'e equation.

Before going to the next section, we want to point out that even though the previous systems actually generalize the Lam\'e equation, the solutions of any of them remain biharmonic functions, as in classical linear elasticity theory.
\section{Preliminaries}
First we recall some definitions and basic properties of a Clifford algebra.

Let $e_1,e_2,e_3$ be an orthonormal basis of $\R^3$. Let $\R_{0,3}$ be real Clifford algebra constructed over $\R^3$. The basic multiplication rules are governed by
\[
e_i^2=-1,\,e_ie_{j}=-e_{j}e_i,\,i,j=1,2,3,\,i<j.
\] 
Any element $a\in\R_{0,m}$ may thus be written as $a=\sum_{A} a_A e_A$, $a_A\in\R$, where $e_A:=e_{i_1}\cdots e_{i_k}$ with $A=\{i_1,\dots,i_k\}\subset\{1,\dots,m\}$ is such that $i_1<\cdots <i_k$. Additionally, one puts $e_\emptyset=1$.  

An element $a\in\R_{0,3}$  can be alternatively written as  
\begin{equation}\label{canonic}
a=[a]_{0}+[a]_{1}+[a]_{2}+[a]_{3},
\end{equation}
where $[\,]_{k}$ denotes the projection of $\R_{0,3}$ onto the subspace $\R_{0,3}^{(k)}$ of $k$-vectors defined by
\[\R_{0,3}^{(k)}={\mbox{span}}_{\R}(e_{A}:\;|A|=k).\]
  
The conjugation in $\R_{0,3}$ is defined as the anti-involution $a\mapsto\overline{a}$ for which $\overline{e_i}=-e_i$. A norm $\|.\|$ on $\R_{0,3}$ is defined by $\|a\|^2=Sc[a\overline{a}]$ for $a\in\R_{0,3}$. We remark that for $\ux\in\R^3$ we have $\|\ux\|=|\ux|$, the usual Euclidean norm.

We will consider functions defined on subsets of $\R^{3}$ and taking values in $\R_{0,3}$. Those functions might be written as $f=\sum_{A} f_A e_A$, where $f_A$ are $\R$-valued functions. The notions of continuity, differentiability and integrability of a $\R_{0,3}$-valued function $f$ have the usual component-wise meaning. In particular, the spaces of all $k$-time continuous differentiable and $p$-integrable functions are denoted by $C^k(\E)$ and $L^p(\E)$ respectively, where $\E$ can be any suitable subset of $\R^{3}$.

The so-called Dirac operator $\upa$ is defined by
\[
\upa:=e_1{\frac{\partial}{\partial x_1}}+e_2{\frac{\partial}{\partial x_2}}+e_3{\frac{\partial}{\partial x_3}}.
\]
An $\R_{0,3}$-valued function $f$, defined and differentiable in an open region $\Omega$ of $\R^{3}$, is called left  monogenic (right monogenic) in $\Omega$ if $\upa f = 0$ ($f \upa= 0$) in $\Omega$. Functions that are both left and right monogenic are called two-sided monogenic.

More generally, for fixed orthonormal base $\psi:=\{\psi^1,\psi^2,\psi^3\}$ in $\R^3$ (structural set) we introduce the so-called $\psi$-hyperholomorphic functions (left or right respectively), which belong to $\ker[\pD(\cdot)]$ or $\ker[(\cdot)\pD]$, where
\begin{equation}\label{DiracPsiLeft}
\pD:=\psi^1{\frac{\partial}{\partial x_1}}+\psi^2{\frac{\partial}{\partial x_2}}+\psi^3{\frac{\partial}{\partial x_3}}.
\end{equation}

For further use we introduce for an open set $\Omega\subset\R^3$, the following subclasses of $\R_{0,3}$-valued functions
\[
\I_{\phi,\psi}(\Omega)=\{u\in C^2(\Omega):\,\hD u\pD=0\},
\]
\[
\cH_{\phi,\psi}(\Omega)=\{u\in C^2(\Omega):\,\hD\pD u=0\}.
\]
It is wise to note here that for $\phi=\psi$, the class $\cH_{\phi,\psi}(\Omega)$ coincides with the space $\cH(\Omega)$ of harmonic functions in $\Omega$. On the other hand and no less important, we note that in case of being $\phi=\psi=\{e_1,e_2,e_3\}$, the class $\I_{\phi,\psi}(\Omega)$ becomes the space $\I(\Omega)$ of inframonogenic functions introduced in \cite{MPS1,MPS2} and studied more extensively in \cite{MAB1,MAB2,MAB3}. The above is reason enough to name the elements of $\I_{\phi,\psi}$ as $(\phi,\psi)$-inframonogenic functions while the elements of $\cH_{\phi,\psi}$ as $(\phi,\psi)$-harmonic functions. For vector-valued functions we will use the alternative notations $\uI_{\phi,\psi}(\Omega)$ and $\ucH_{\phi,\psi}(\Omega)$.
\section{Auxiliary results}
In order to prove the main results, we will establish some auxiliary results which are provided in this section.
\begin{proposition}\label{prop1}
An $\R_{0,3}$-valued function $f$ is $(\psi,\psi)$-inframonogenic in  $\Omega\subset\R^3$ if and only if each $k$-vector valued function $[f]_k$, $0\le k\le 3$,  is $(\psi,\psi)$-inframonogenic there.
\end{proposition}
\pf
The proof is adapted from \cite{MAB1}. The only new ingredient to use is the representation of $\psi^i$ ($i=1,2,3$) by
\[\psi^i=\sum_{\overset{k=1}{}}^{3}\,\psi_k^ie_k,\] 
where $\psi_k^i\in\mathbb{R}$.

After this, the proof follows very similar lines of the proof of Proposition 1 in \cite{MAB1} and will be omitted.\qed

In contrast to the particular case $\phi=\psi$, the above nice property is no longer valid in general.  As a simple counterexample consider the function given by
\begin{equation}\label{ce2}
g(x)=\frac{1}{2}x_1^2+\frac{1}{2}x_2^2+\sqrt{2}x_1x_2+(e_3e_1-1)x_3^2.
\end{equation}	
Let be $\phi=\{e_1,e_3,e_2\}$ and $\psi=\{\frac{\sqrt{2}}{2}(e_1+e_3),\frac{\sqrt{2}}{2}(e_1-e_3),e_2\}$ two structural sets.

On the one hand, we have 	
\begin{align*}
	\hD g\pD &=e_1\left[\frac{\sqrt{2}}{2}(e_1+e_3)\right]+e_3\left[\frac{\sqrt{2}}{2}(e_1-e_3)\right]+e_1\sqrt{2}\left[\frac{\sqrt{2}}{2}(e_1-e_3)\right]\\&\;\;\;\;+e_3\sqrt{2}\left[\frac{\sqrt{2}}{2}(e_1+e_3)\right]+e_2[2e_3e_1-2]e_2\\
	&=-\frac{\sqrt{2}}{2}+\frac{\sqrt{2}}{2}e_1e_3+\frac{\sqrt{2}}{2}e_3e_1+\frac{\sqrt{2}}{2}-1-e_1e_3+e_3e_1-1-2e_3e_1+2\\
	&=0,
	\end{align*}
but, on the other
\begin{align*}
\hD [g]_0\pD&=\hD\left[\frac{1}{2}x_1^2+\frac{1}{2}x_2^2-x_3^2+\sqrt{2}x_1x_2\right]\pD
\\&=-\frac{\sqrt{2}}{2}+\frac{\sqrt{2}}{2}e_1e_3+\frac{\sqrt{2}}{2}e_3e_1+\frac{\sqrt{2}}{2}-1-e_1e_3+e_3e_1-1+2
\\&=2e_3e_1\not=0,\\
\hD[g]_1\pD&=0,
\\\hD[g]_2\pD&=\hD[x_3^2e_3e_1]\pD\\&=-2e_3e_1\not=0,
\\\hD[g]_3\pD&=0.
\end{align*}
Let as before $\psi=\{\psi^1,\psi^2,\psi^3\}$ and $\phi=\{\phi^1,\phi^2,\phi^3\}$ be two structural sets in $\R^3$. With the notations
$$\nu(f)=\sum_{\overset{i=1}{}}^{3}\,\psi^i f \psi^i ,\;\;\;\ \omega(f)=\sum_{\overset{i=1}{}}^{3}\,\phi^i f \psi^i,\;\;\;\ \widetilde{\omega}(f)=\sum_{\overset{i=1}{}}^{3}\,\psi^i f \phi^i,$$
we have
\begin{lemma}\label{l1}
Let $f:\R^3\rightarrow\mathbb{R}_{0,3}$ and $\ux_\psi=\sum_{\overset{i=1}{}}^{3}\,\psi^i x_i$. Then,
\begin{itemize}
\item[(1)] $\pD(f\ux_\psi)=(\pD f)\underline{x}_\psi+\nu(f),\;\;\;\;(\underline{x}_\psi f)\pD=\ux_\psi(f\pD)+\nu(f)$
\item[(2)] $\pD[\nu(f)]=-2f\pD-\nu(\pD f),\;\;\;\;[\nu(f)]\pD=-2\pD f-\nu(f\pD)$
\item[(3)] $\pD[\nu(f)]\pD=\nu(\pD f \pD),\;\;\;\;[\nu(f)]\pD\pD=-\Delta \nu(f)=\nu(\pD\pD f)=-\nu(\Delta f)$
\item[(4)] $\hD[\omega(f)]\pD=\omega(\hD f\pD),\;\;\;\;[\omega(f)]\pD\pD=-\Delta \omega(f)=\omega(\pD\pD f)=-\omega(\Delta f)$
\item[(5)] $\nu(\vec{u})=\vec{u}$
\item[(6)] $\hD\omega(f)=-2f\pD-\omega(\hD f),\;\;\;\;\;\omega(f)\pD=-2\hD f-\omega(f\pD)$
\item[(7)] $\hD(f\underline{x}_\psi)=(\hD f)\underline{x}_\psi+\omega(f),\;\;\;\;\;(\underline{x}_\psi f)\hD=\underline{x}_\psi(f\hD)+\widetilde{\omega}(f)$
\item[(8)] $\pD\hD[\omega(f)]-\omega(\hD\pD f)=\hD[\omega(\pD f)]-\pD[\omega(\hD f)]$
%\item[(9)] $\pD\hD[\omega(f)]=-2\pD f\pD-\pD[\omega(\hD f)]$
%\item[(10)] $\omega(\hD\pD f)=-2\pD f\pD-\hD[\omega(\pD f)]$
\end{itemize}
\end{lemma}
\pf For the sake of brevity we only include the proofs of $(6)$ and $(7)$. The remaining statements follow by similar arguments.

{\it Proof of (6):}
\begin{equation*}
\begin{split}
\hD\omega(f)&=\displaystyle\sum_{\overset{1\leq i,j\leq 3}{}}\,\phi^i\phi^j(\partial_{x_i}f)\psi^j\\
&=\displaystyle\sum_{\overset{i=j}{1\leq i,j\leq 3}}\,\phi^i\phi^j(\partial_{x_i}f)\psi^j+\displaystyle\sum_{\overset{i\not=j}{1\leq i,j\leq 3}}\,\phi^i\phi^j(\partial_{x_i}f)\psi^j\\
&=-\displaystyle\sum_{\overset{i=1}{}}^{3}\,(\partial_{x_i}f)\psi^i-\displaystyle\sum_{\overset{i\not=j}{1\leq i,j\leq 3}}\,\phi^j\phi^i(\partial_{x_i}f)\psi^j\\
&=-\displaystyle\sum_{\overset{i=1}{}}^{3}\,(\partial_{x_i}f)\psi^i-(\displaystyle\sum_{\overset{1\leq i,j\leq 3}{}}\,\phi^j\phi^i(\partial_{x_i}f)\psi^j+\displaystyle\sum_{\overset{i=1}{}}^{3}\,(\partial_{x_i}f)\psi^i)\\
&=-2\displaystyle\sum_{\overset{i=1}{}}^{3}\,(\partial_{x_i}f)\psi^i-\displaystyle\sum_{\overset{1\leq i,j\leq 3}{}}\,\phi^j\phi^i(\partial_{x_i}f)\psi^j\\
&=-2f\pD-\omega(\hD f).
\end{split}
\end{equation*}
{\it Proof of (7):}
\begin{align*}
\hD(f\ux_\psi)&=\displaystyle \sum_{\overset{i=1}{}}^{3}\, \phi^i\frac{\partial(f\ux_\psi)}{\partial x_i}&(\ux_\psi f)\hD&=\displaystyle \sum_{\overset{i=1}{}}^{3}\,\frac{\partial(\ux_\psi f)}{\partial x_i}\phi^i\\
&=\sum_{\overset{i=1}{}}^{3}\,\phi^i\left(\frac{\partial f}{\partial x_i}\ux_\psi+f\psi^i\right) &&=\sum_{\overset{i=1}{}}^{3}\,\left(\psi^if+\ux_\psi\frac{\partial f}{\partial x_i}\right)\phi^i\\
&=\displaystyle \left(\sum_{\overset{i=1}{}}^{3}\,\phi^i\frac{\partial f}{\partial x_i}\right)\ux_\psi+\sum_{\overset{i=1}{}}^{3}\,\phi^if\psi^i &&=\displaystyle \ux_\psi\left(\sum_{\overset{i=1}{}}^{3}\,\frac{\partial f}{\partial x_i}\phi^i\right)+\sum_{\overset{i=1}{}}^{3}\,\psi^if\phi^i\\
&=(\hD f)\ux_\psi+\omega(f).&&=\ux_\psi(f\hD)+\widetilde{\omega}(f).
\end{align*}

As was explicitly mentioned at the end of the introduction, the solutions of \eqref{lamephipsi} are biharmonic functions. The following stronger result is in fact true.
\begin{proposition}
If $\vec{u}\in C^3(\Omega)$ satisfies in $\Omega$ the generalized Lam\'e-Navier system \eqref{lamephipsi}, then $\pD^3\vec{u}=0$ in $\Omega$.
\end{proposition} 
\pf
On applying $\hD$ to both sides of \eqref{lamephipsi} yields 
\begin{equation*}
\alpha \hD\hD\vec{u}\pD+\beta\hD\hD\pD\vec{u}=0,
\end{equation*}
and hence
\begin{equation*}
\alpha \vec{u}\pD\hD\hD+\beta\pD\pD\pD\vec{u}=0.
\end{equation*}
Since  $\vec{u}\pD\hD=-\frac{\alpha}{\beta}\pD\vec{u}\hD$, it follows that
\begin{equation*}
-\frac{\alpha^2}{\beta}{\pD}\vec{u}\hD\hD+\beta\pD\pD\pD\vec{u}=0
\end{equation*}
or equivalently
\begin{equation*}
{\left(\beta-\frac{\alpha^2}{\beta}\right)}\pD\pD\pD\vec{u}=0.
\end{equation*}

From this, we conclude that $\pD^3\vec{u}=0$, as otherwise would be $\alpha=\beta$ or $\alpha=-\beta$ and one is led to a contradiction with the original assumptions on the Lam\'e constants $\mu$, $\lambda$.
\qed
\section{Additive decomposition of the generalized Lam\'e-Navier solutions}
In this section our main results are stated and proved. We start by a rather simple generalization of \cite[Theorem 3.1]{MAB3}.

\begin{theorem}\label{T1}
If a vector field $\vec{u}$ satisfies in $\Omega\subset\R^3$ the generalized Lam\'e-Navier system \eqref{lamepsi}, then it admits in $\Omega$ the decomposition
\[\vec{u}=\vec{h}+\vec{i},\]
where $\vec{h}\in\ucH(\Omega)$ and $\vec{i}\in\uI_{\psi,\psi}(\Omega)$. Moreover, this representation is unique up to a vector field in $\ucH(\Omega)\cap\uI_{\psi,\psi}(\Omega)$. 
\end{theorem}
\pf  Let $g=\alpha\vec{u}\pD+\beta \pD\vec{u}$, $\vec{u}$ satisfying \eqref{lamepsi}. Clearly $g$ is a $\R_{0,3}$-valued (left) $\psi$-hyperholomorphic function in $\Omega$. Moreover, in virtue of Lemma \ref{l1} $(1)$  we have $\pD(g\ux_\psi)=\nu(g)$, which by Lemma \ref{l1} $(2)$-$(5)$ yields 
$$\pD(g\ux_\psi)\pD=-\nu(g\pD)=-g\pD=\left(\frac{\alpha^2}{\beta}-\beta\right)\pD\vec{u}\pD$$and
$$\pD\pD(g\ux_\psi)=-2g\pD=2\left(\frac{\beta^2}{\alpha}-\alpha\right)\pD\pD\vec{u}.$$
Equivalently:
\begin{equation}\label{1}
\pD\left[g\ux_\psi-\left(\frac{\alpha^2}{\beta}-\beta\right)\vec{u}\right]\pD=0
\end{equation}
and
\begin{equation}\label{2}
\pD\pD\left[g\ux_\psi-2\left(\frac{\beta^2}{\alpha}-\alpha\right)\vec{u}\right]=0.
\end{equation}

Let be $I:=g\ux_\psi-(\frac{\alpha^2}{\beta}-\beta)\vec{u}$ and $H:=g\ux_\psi-2(\frac{\beta^2}{\alpha}-\alpha)\vec{u}$. Of course, since \eqref{1}-\eqref{2} $I\in\I_{\psi,\psi}(\Omega)$, $H\in\mathcal{H}(\Omega)$ and
$$\left(\frac{\alpha^2}{\beta}-\beta-\frac{2\beta^2}{\alpha}+2\alpha\right)\vec{u}=H-I.$$
Our next task is to prove that $\frac{\alpha^2}{\beta}-\beta-\frac{2\beta^2}{\alpha}+2\alpha\not=0$, or equivalently, that 
\[
(\alpha+2\beta)(\alpha^2-\beta^2)\not=0.
\]
Indeed, if $\alpha+2\beta=0$ we would have $\frac{3\lambda}{\mu}=-7$ and then $\frac{\lambda}{\mu}<-\frac{2}{3}$, which contradicts the initial assumption on the Lam\'e coefficients $\lambda,\mu$. In a similar way the supposition $\alpha^2-\beta^2=0$ leads to a contradiction.

Therefore, we have 
\begin{equation}\label{decompGen}
\vec{u}=h+i, 
\end{equation}
where 
\[
h=\left(\frac{\alpha^2}{\beta}-\beta-\frac{2\beta^2}{\alpha}+2\alpha\right)^{-1}H,\,i=-\left(\frac{\alpha^2}{\beta}-\beta-\frac{2\beta^2}{\alpha}+2\alpha\right)^{-1}I.
\]
Since $h\in{\mathcal H}(\Omega)$ and $i\in\mathcal{I}_{\psi,\psi}(\Omega)$, the desired representation easily follows from Proposition \ref{prop1} and taking the $1$-vector part in both sides of \eqref{decompGen}. 

The proof of the second part is obvious. Indeed, assume that $\vec{u}$, being a solution of \eqref{Lame} admits two different representations, say,  
\[\vec{u}=\vec{h}_1+\vec{i}_1,\,\vec{u}=\vec{h}_2+\vec{i}_2,\]
where $\vec{h}_{1},\vec{h}_{2}\in\underline{{\mathcal H}}(\Omega)$ and $\vec{i}_{1},\vec{i}_{2}\in\underline{\mathcal{I}}_{\psi,\psi}(\Omega)$. 

Then by subtracting both representations we obtain that $\vec{h}_1-\vec{h}_2=\vec{i}_2-\vec{i}_1$ are simultaneously harmonic and $(\psi,\psi)$-inframonogenic. \qed
From now on we will be concerned with the much more general Lam\'e-Navier system \eqref{lamephipsi}. Since Proposition 1 is no longer available in this general situation, the next decomposition theorems involve $\R_{0,3}$-valued functions rather than simply vector-valued ones. We start with a technical lemma, whose proof is a matter of direct calculations.
\begin{lemma}\label{l2}
Let $\vec{u}$ satisfy \eqref{lamephipsi} in $\Omega\subset\mathbb{R}^3$ and put $g=\alpha\vec{u}\pD+\beta \pD\vec{u}$. Then
\begin{equation}\label{aux1}
\hD\pD(g\ux_\psi)=(\hD\pD g)\ux_\psi-\frac{\alpha}{\beta}\hD(g\ux_\psi)\pD+\left(\frac{\alpha^2}{\beta}-\beta\right)\omega(\Delta\vec{u})+2\left(\frac{\beta^2}{\alpha}-\alpha\right)\hD\pD\vec{u},
\end{equation}
and
\begin{equation}\label{aux2}
\hD\pD\left[g\ux_\psi-\frac{\alpha}{\beta}\overline{g}\ux_\psi-\left(\frac{\alpha^2}{\beta}-\beta+\frac{2\beta^2}{\alpha}-2\alpha\right)\vec{u}\right]=(\hD\pD g)\ux_\psi+\left(\frac{\alpha^2}{\beta}-\beta\right)\omega(\Delta\vec{u}).
\end{equation}
\end{lemma} 
 Here is a generalization of Theorem \ref{T1}.
\begin{theorem}\label{T2}
Let $\vec{u}$ satisfy \eqref{lamephipsi} in $\Omega\subset\mathbb{R}^3$. If $\vec{u}$ is harmonic and $(\psi,\psi)$-inframonogenic in $\Omega$, then it admits there the splitting
\[\vec{u}=h+i,\]
where $h\in\mathcal{H}_{\phi,\psi}(\Omega)$ and $i\in\mathcal{I}_{\phi,\psi}(\Omega)$. Moreover, this representation is unique up to an element in $\mathcal{H}_{\phi,\psi}(\Omega)\cap\mathcal{I}_{\phi,\psi}(\Omega)$.  
\end{theorem}
\pf
Once again, let $g=\alpha\vec{u}\pD+\beta \pD\vec{u}$. Under the assumptions stated above, we have $\hD g=0$, $\pD g=0$ and $g\pD =0$. 

Lemma \ref{l1} $(6)$-$(7)$ now yields 
\begin{equation*}\label{omegag}
\hD(g\ux_\psi)\pD=\omega(g)\pD=-2\hD g-\omega(g\pD)=0.
\end{equation*}
Consequently, by \eqref{aux1}  we have
\[\hD\pD(g\ux_\psi)=2\left(\frac{\beta^2}{\alpha}-\alpha\right)\hD\pD\vec{u},\] 
or equivalently
$$\hD\pD\left[g\ux_\psi-2\left(\frac{\beta^2}{\alpha}-\alpha\right)\vec{u}\right]=0.$$
Since $\alpha-\frac{\beta^2}{\alpha}\not=0$, the proof is completed after  taking
\[
h:=\left(2\alpha-\frac{2\beta^2}{\alpha}\right)^{-1}\left[g\ux_\psi-2\left(\frac{\beta^2}{\alpha}-\alpha\right)\vec{u}\right],\,i:=-\left(2\alpha-\frac{2\beta^2}{\alpha}\right)^{-1}g\ux_\psi.
\]
The uniqueness is obvious and its proof will be omitted.
\qed
Now, we will show how to deal without imposing any assumption of inframonogenicity. As we will see, a subtle change is needed in replacing the space $\mathcal{I}_{\phi,\psi}(\Omega)$ of the previous theorem by $\mathcal{I}_{\psi,\phi}(\Omega)$.
\begin{theorem}
Let $\vec{u}$ satisfy \eqref{lamephipsi} in $\Omega\subset\mathbb{R}^3$. If $\vec{u}$ is harmonic in $\Omega$, then it admits the decomposition
\[\vec{u}=h+i^{*},\]
where $h\in\cH_{\phi,\psi}(\Omega)$ and $i^{*}\in\I_{\psi,\phi}(\Omega)$. Moreover, this representation is unique up to an element in $\mathcal{H}_{\phi,\psi}(\Omega)\cap\mathcal{I}_{\psi,\phi}(\Omega)$.  
\end{theorem}
\pf
Let $g=\alpha\vec{u}\pD+\beta \pD\vec{u}$ and $\og=\alpha\pD\vec{u}+\beta\vec{u}\pD$. Then 
\[\hD\pD g=\hD\overline{g}\pD =\left(\beta-\frac{\alpha^2}{\beta}\right)\hD\vec{u}\pD\pD=0,\]
since $\vec{u}$ is harmonic.
 
By applying \eqref{aux2} we obtain 
\begin{equation}\label{eq1}
\hD\pD\left[\left(g-\frac{\alpha}{\beta}\og\right)\ux_\psi-\left(\frac{\alpha^2}{\beta}-\beta+\frac{2\beta^2}{\alpha}-2\alpha\right)\vec{u}\right]=0.
\end{equation}
On the other hand, the relation $(g-\frac{\alpha}{\beta}\og)\ux_\psi=(\beta-\frac{\alpha^2}{\beta})(\pD\vec{u})\ux_\psi$ implies 
\begin{equation*}
\begin{split}
\pD\left[\left(g-\frac{\alpha}{\beta}\og\right)\ux_\psi\right]&=\pD\left[\left(\beta-\frac{\alpha^2}{\beta}\right)(\pD\vec{u})\ux_\psi\right]\\
&=\left(\beta-\frac{\alpha^2}{\beta}\right)[(\pD\pD\vec{u})\ux_\psi+\nu(\pD\vec{u})]\\
&=\left(\beta-\frac{\alpha^2}{\beta}\right)(-2\vec{u} \pD-\pD\vec{u}).
\end{split}
\end{equation*}
Consequently
$$\pD\left[\left(g-\frac{\alpha}{\beta}\og\right)\ux_\psi\right]\hD=\left(\beta-\frac{\alpha^2}{\beta}\right)(-2\vec{u} \pD\hD-\pD\vec{u}\hD)=\left(\beta-\frac{\alpha^2}{\beta}\right)\left(\frac{2\alpha}{\beta}-1\right)\pD\vec{u}\hD$$
or equivalently
\begin{equation}\label{eq2}
\pD\left[\left(g-\frac{\alpha}{\beta}\og\right)\ux_\psi-\left(2\alpha-\beta-\frac{2\alpha^3}{\beta^2}+\frac{\alpha^2}{\beta}\right)\vec{u}\right]\hD=0.
\end{equation}

Next let be 
\[I^{*}=\left(g-\frac{\alpha}{\beta}\og\right)\ux_\psi-\left(2\alpha-\beta-\frac{2\alpha^3}{\beta^2}+\frac{\alpha^2}{\beta}\right)\vec{u}\]
and 
\[H=\left(g-\frac{\alpha}{\beta}\og\right)\ux_\psi-\left(2\alpha-\beta-\frac{2\alpha^3}{\beta^2}+\frac{\alpha^2}{\beta}\right)\vec{u}+\left(4\alpha-\frac{2\alpha^3}{\beta^2}-\frac{2\beta^2}{\alpha}\right)\vec{u}.
\] 
The proof is now easily completed from \eqref{eq1}-\eqref{eq2} by choosing $h=(4\alpha-\frac{2\alpha^3}{\beta^2}-\frac{2\beta^2}{\alpha})^{-1}H$ and $i^{*}=-(4\alpha-\frac{2\alpha^3}{\beta^2}-\frac{2\beta^2}{\alpha})^{-1}I^{*}$. The factor $4\alpha-\frac{2\alpha^3}{\beta^2}-\frac{2\beta^2}{\alpha}$ is not $0$, since one would otherwise obtain a contradiction with the assumptions on the Lam\'e parameters.

Once again, the proof of uniqueness is straightforward and will be omitted. \qed

Similarly we have a corresponding theorem without recourse to the assumption of harmonicity.
\begin{theorem}\label{T3}
If a $(\psi,\psi)$-inframonogenic vector field $\vec{u}$ satisfies in $\Omega\subset\mathbb{R}^3$ the generalized Lam\'e-Navier system \eqref{lamephipsi}, then it admits the representation
$$\vec{u}=h+i^{*},$$
where $h\in\mathcal{H}_{\phi,\psi}(\Omega)$ and $i^{*}\in\mathcal{I}_{\psi,\phi}(\Omega)$.  Moreover, this representation is unique up to an element in $\mathcal{H}_{\phi,\psi}(\Omega)\cap\mathcal{I}_{\psi,\phi}(\Omega)$.    
\end{theorem}
\section{Construction of solutions}
In this section we give a direct method for constructing solutions of \eqref{lamephipsi} from harmonic and/or inframonogenic functions. 
\begin{theorem}
If $u$ is harmonic or $(\phi,\psi)$-inframonogenic in $\Omega$, then 
\[w=u\pD-\frac{\beta}{\alpha}\pD u\] 
satisfies \eqref{lamephipsi}. 
\end{theorem}
\pf
Indeed, we have
\begin{equation*}
\begin{split}
\alpha[\hD w\pD]+\beta[\hD\pD w]\\=\alpha \hD u\pD\pD-\beta \hD\pD u\pD+\beta \hD\pD u\pD-\frac{\beta^2}{\alpha}\hD\pD\pD u\\=
\left(\alpha-\frac{\beta^2}{\alpha}\right)\hD u\pD\pD=0,
\end{split}
\end{equation*}
which is due to the fact that $u$ is harmonic or $(\phi,\psi)$-inframonogenic in $\Omega$.
\qed
Notice that the above solution is in general $\R_{0,3}$-valued but it becomes vector-valued if $u$ is a scalar function, as is easy to check. The same fact is valid in the following theorem.
\begin{theorem}\label{T4}
If $u$ is $(\phi,\psi)$-harmonic or $(\psi,\psi)$-inframonogenic in $\Omega$ then
$$\widetilde{w}=u\pD-\frac{\alpha}{\beta}\pD u$$
satisfies \eqref{lamephipsi}. 
\end{theorem}
The following result shows how to a given harmonic and $(\phi,\psi)$-harmonic vector field $\vec{h}$, corresponds a sort of $(\phi,\psi)$-inframonogenic conjugate function $i$ such that $\vec{h}+i$ represents a solution of \eqref{lamephipsi}.
\begin{theorem}
Let $\vec{h}\in\underline{\mathcal{H}}(\Omega)\cap\underline{\mathcal{H}}_{\phi,\psi}(\Omega)$ and suppose $\omega(\pD\vec{h})=-\hD\vec{h}$. Then, there exists a function $i\in\mathcal{I}_{\phi,\psi}(\Omega)$ such that $\vec{h}+i$ solves \eqref{lamephipsi}. Moreover, $i$ may be represented as $i=\frac{\alpha}{2\beta}[\vec{h}+(\pD h)\ux_\psi]$.   
\end{theorem}
\pf
A direct calculation gives
$$\hD i\pD=\frac{\alpha}{2\beta}[\hD\vec{h}+(\hD\pD\vec{h})\ux_\psi+\omega(\pD\vec{h})]\pD=\frac{\alpha}{2\beta}[\hD\vec{h}\pD +\omega(\pD\vec{h})\pD]=0.$$
On the other hand,
\begin{equation*}
\begin{split}
\alpha[\hD(\vec{h}+i)\pD]+\beta[\hD\pD(\vec{h}+i)]&=\alpha \hD\vec{h}\pD+\beta \hD\pD i\\
&=\alpha \hD\vec{h}\pD+\frac{\alpha}{2}\hD\pD[(\pD\vec{h})\ux_\psi]\\
&=\alpha \hD\vec{h}\pD+\frac{\alpha}{2}\hD\nu(\pD\vec{h})\\
&=\alpha \hD\vec{h}\pD+\frac{\alpha}{2}\hD\{-2\vec{h}\pD-\pD[\nu(\vec{h})]\}\\
&=-\frac{\alpha}{2}\hD\pD[\nu(\vec{h})]\\
&=-\frac{\alpha}{2}\hD\pD\vec{h}\\
&=0.
\end{split}
\end{equation*}
And we are done. \qed

It is worth noting that the above proof strongly depended on the assumption that $\vec{u}$ is a vector-valued function.

The following result is also obtained in a similar way:
\begin{theorem}
Let $\vec{i}\in\underline{\mathcal{I}}_{\psi,\psi}(\Omega)\cap\underline{\mathcal{I}}_{\phi,\psi}(\Omega)$ and suppose $\hD\omega(\vec{i})=\pD\vec{i}$. Then, there exists a function $h\in\mathcal{H}_{\phi,\psi}(\Omega)$ such that $h+\vec{i}$ solves \eqref{lamephipsi}. Moreover, $h$ may be represented as $h=\frac{\beta}{\alpha}[2\vec{i}+(\vec{i}\pD)\ux_\psi]$.
\end{theorem}
At the end of the paper (see Appendix) a function in MATLAB is provided for performing computations using the Clifford algebra reformulation of both classical and generalized Lam\'e-Navier systems. Moreover, the implemented MATLAB function is used to verify some algebraic results obtained in the paper.

As an example, consider the generalized Lam\'e-Navier system
\begin{equation}\label{ejemplo}
\alpha[\hD\vec{u}\upa]+\beta[\hD\upa\vec{u}]=0,
\end{equation}
where $\alpha=0.1$, $\beta=0.2$ and $\phi=\{-e_1,e_2,e_3\}$.

Applying the function {\bf Lame} to the harmonic vector field
\begin{equation}\label{study}
\vec{u}=x_1x_2e_1+(-2x_1^2-3x_2^2+5x_3^2)e_2+x_3e_3,
\end{equation}
we verify that it is a solution of \eqref{ejemplo}.

On the other hand, after applying the procedure carried out in Theorem 3, we arrive to the decomposition
\[
\vec{u}(\ux)=h+i^*,
\] 
where
\begin{align*}
h(\ux)&=-\frac{20}{9}[(1.05x_2-0.15)x_1e_1+(0.15x_1^2+2.10x_2^2-0.75x_3^2-0.15x_2)e_2
\\&\;\;\;\;-(0.60+0.75x_2)x_3e_3-2.25x_1x_3e_1e_2e_3],\\
i^*(\ux)&=\frac{20}{9}[(1.50x_2-0.15)x_1e_1+(0.75x_2^2+1.50x_3^2-0.75x_1^2-0.15x_2)e_2
\\&\;\;\;\;-(0.15+0.75x_2)x_3e_3-2.25x_1x_3e_1e_2e_3],
\end{align*}     
satisfy $\hD\upa h=0$ and $\upa i^*\hD=0$, respectively.

Finally, we remark that such a vector field \eqref{study} is also a particular solution of the inhomogeneous classical Lam\'e system
\begin{equation}\label{inh}
\alpha[\upa\vec{u}\upa]+\beta[\upa^2\vec{u}]=e_2.
\end{equation}
This fact suggests the idea that, if structural sets are conveniently chosen, some kinds of classical inhomogeneous Lam\'e systems (in presence of a constant volume force) may be rewritten as a homogeneous one. But we will not develop this point here.  
\section{Appendix: MATLAB implementation}
{\footnotesize{
\begin{verbatim}
function [DphiDpsif,DpsiDphif,DphifDpsi,DpsifDphi,
DpsifDpsi,DphifDphi,DfD,D2f,Lcf,Lcphif,Lcpsif,Lgf,Lgif]= Lame(Phi1,
Phi2,Phi3,Psi1,Psi2,Psi3,F,Cl)
syms 'x1' 'x2' 'x3';
phi11=Phi1(1);phi12=Phi1(2);phi13=Phi1(3);phi21=Phi2(1);phi22=Phi2(2);
phi23=Phi2(3);phi31=Phi3(1);phi32=Phi3(2);phi33=Phi3(3);psi11=Psi1(1);
psi12=Psi1(2);psi13=Psi1(3);psi21=Psi2(1);psi22=Psi2(2);psi23=Psi2(3);
psi31=Psi3(1);psi32=Psi3(2);psi33=Psi3(3);
alpha=Cl(1);beta=Cl(2);
alphaC=Clifford([0 3 0],[alpha 0 0 0 0 0 0 0]);
betaC=Clifford([0 3 0],[beta 0 0 0 0 0 0 0]); 
if beta-alpha>0 && 7*alpha>beta;
        disp('Coefficients meet Lame`s restrictions')
else
        disp('Coefficients do not meet Lame`s restrictions')
end
k0=F(1);k1=F(2);k2=F(3);k3=F(4);k4=F(5);k5=F(6);k6=F(7);k7=F(8);
dk0x1=diff(k0,x1);dk1x1=diff(k1,x1);dk2x1=diff(k2,x1);
dk3x1=diff(k3,x1);dk4x1=diff(k4,x1);dk5x1=diff(k5,x1);
dk6x1=diff(k6,x1);dk7x1=diff(k7,x1);dk0x2=diff(k0,x2);
dk1x2=diff(k1,x2);dk2x2=diff(k2,x2);dk3x2=diff(k3,x2);
dk4x2=diff(k4,x2);dk5x2=diff(k5,x2);dk6x2=diff(k6,x2);
dk7x2=diff(k7,x2);dk0x3=diff(k0,x3);dk1x3=diff(k1,x3);
dk2x3=diff(k2,x3);dk3x3=diff(k3,x3);dk4x3=diff(k4,x3);
dk5x3=diff(k5,x3);dk6x3=diff(k6,x3);dk7x3=diff(k7,x3);
dfx1=Clifford([0 3 0],[dk0x1,dk1x1,dk2x1,dk3x1,dk4x1,dk5x1,dk6x1,
dk7x1]);
dfx2=Clifford([0 3 0],[dk0x2,dk1x2,dk2x2,dk3x2,dk4x2,dk5x2,dk6x2,
dk7x2]);
dfx3=Clifford([0 3 0],[dk0x3,dk1x3,dk2x3,dk3x3,dk4x3,dk5x3,dk6x3,
dk7x3]);
phi1=Clifford([0 3 0],[0 phi11 phi12 phi13 0 0 0 0]);
phi2=Clifford([0 3 0],[0 phi21 phi22 phi23 0 0 0 0]);
phi3=Clifford([0 3 0],[0 phi31 phi32 phi33 0 0 0 0]);
psi1=Clifford([0 3 0],[0 psi11 psi12 psi13 0 0 0 0]);
psi2=Clifford([0 3 0],[0 psi21 psi22 psi23 0 0 0 0]);
psi3=Clifford([0 3 0],[0 psi31 psi32 psi33 0 0 0 0]);
dk0x11=diff(dk0x1,x1);dk0x22=diff(dk0x2,x2);dk0x23=diff(dk0x2,x3);
dk0x33=diff(dk0x3,x3);dk1x11=diff(dk1x1,x1); dk1x22=diff(dk1x2,x2);
dk1x23=diff(dk1x2,x3);dk1x33=diff(dk1x3,x3);dk2x11=diff(dk2x1,x1);
dk2x22=diff(dk2x2,x2);dk2x23=diff(dk2x2,x3);dk2x33=diff(dk2x3,x3);
dk3x11=diff(dk3x1,x1);dk3x22=diff(dk3x2,x2);dk3x23=diff(dk3x2,x3);
dk3x33=diff(dk3x3,x3);dk4x11=diff(dk4x1,x1); dk4x22=diff(dk4x2,x2);
dk4x23=diff(dk4x2,x3);dk4x33=diff(dk4x3,x3);dk5x11=diff(dk5x1,x1);
dk5x22=diff(dk5x2,x2);dk5x23=diff(dk5x2,x3);dk5x33=diff(dk5x3,x3);
dk6x11=diff(dk6x1,x1);dk6x22=diff(dk6x2,x2);dk6x23=diff(dk6x2,x3);
dk6x33=diff(dk6x3,x3);dk7x11=diff(dk7x1,x1); dk7x22=diff(dk7x2,x2);
dk7x23=diff(dk7x2,x3);dk7x33=diff(dk7x3,x3);dk0x12=diff(dk0x1,x2);
dk1x12=diff(dk1x1,x2);dk2x12=diff(dk2x1,x2);dk3x12=diff(dk3x1,x2);
dk4x12=diff(dk4x1,x2);dk5x12=diff(dk5x1,x2);dk6x12=diff(dk6x1,x2);
dk7x12=diff(dk7x1,x2);dk0x13=diff(dk0x1,x3);dk1x13=diff(dk1x1,x3);
dk2x13=diff(dk2x1,x3);dk3x13=diff(dk3x1,x3);dk4x13=diff(dk4x1,x3);
dk5x13=diff(dk5x1,x3);dk6x13=diff(dk6x1,x3);dk7x13=diff(dk7x1,x3);
dfx11=Clifford([0 3 0],[dk0x11,dk1x11,dk2x11,dk3x11,dk4x11,dk5x11,
dk6x11,dk7x11]);
dfx22=Clifford([0 3 0],[dk0x22,dk1x22,dk2x22,dk3x22,dk4x22,dk5x22,
dk6x22,dk7x22]);
dfx33=Clifford([0 3 0],[dk0x33,dk1x33,dk2x33,dk3x33,dk4x33,dk5x33,
dk6x33,dk7x33]);
dfx12=Clifford([0 3 0],[dk0x12,dk1x12,dk2x12,dk3x12,dk4x12,dk5x12,
dk6x12,dk7x12]);
dfx13=Clifford([0 3 0],[dk0x13,dk1x13,dk2x13,dk3x13,dk4x13,dk5x13,
dk6x13,dk7x13]);
dfx23=Clifford([0 3 0],[dk0x23,dk1x23,dk2x23,dk3x23,dk4x23,dk5x23,
dk6x23,dk7x23]);
D2f=Clifford([0 3 0],[-1 0 0 0 0 0 0 0])*(dfx11+dfx22+dfx33)
Dphif=phi1*dfx1+phi2*dfx2+phi3*dfx3;fDphi=dfx1*phi1+dfx2*phi2+dfx3*phi3;
Dpsif=psi1*dfx1+psi2*dfx2+psi3*dfx3;fDpsi=dfx1*psi1+dfx2*psi2+dfx3*psi3;
DphiDpsif=phi1*psi1*dfx11+phi1*psi2*dfx12+phi1*psi3*dfx13+phi2*psi1*dfx12
+phi2*psi2*dfx22+phi2*psi3*dfx23+phi3*psi1*dfx13+phi3*psi2*dfx23
+phi3*psi3*dfx33
DpsiDphif=psi1*phi1*dfx11+psi1*phi2*dfx12+psi1*phi3*dfx13
+psi2*phi1*dfx12+psi2*phi2*dfx22+psi2*phi3*dfx23
+psi3*phi1*dfx13+psi3*phi2*dfx23+psi3*phi3*dfx33
DphifDpsi=phi1*dfx11*psi1+phi1*dfx12*psi2+phi1*dfx13*psi3+phi2*dfx12*psi1
+phi2*dfx22*psi2+phi2*dfx23*psi3+phi3*dfx13*psi1+phi3*dfx23*psi2
+phi3*dfx33*psi3
DpsifDphi=psi1*dfx11*phi1+psi1*dfx12*phi2+psi1*dfx13*phi3+psi2*dfx12*phi1
+psi2*dfx22*phi2+psi2*dfx23*phi3+psi3*dfx13*phi1+psi3*dfx23*phi2
+psi3*dfx33*phi3
DphifDphi=phi1*dfx11*phi1+phi1*dfx12*phi2+phi1*dfx13*phi3+phi2*dfx12*phi1
+phi2*dfx22*phi2+phi2*dfx23*phi3+phi3*dfx13*phi1+phi3*dfx23*phi2
+phi3*dfx33*phi3
DpsifDpsi=psi1*dfx11*psi1+psi1*dfx12*psi2+psi1*dfx13*psi3+psi2*dfx12*psi1
+psi2*dfx22*psi2+psi2*dfx23*psi3+psi3*dfx13*psi1+psi3*dfx23*psi2
+psi3*dfx33*psi3
e1=Clifford([0 3 0],[0 1 0 0 0 0 0 0]);
e2=Clifford([0 3 0],[0 0 1 0 0 0 0 0]);
e3=Clifford([0 3 0],[0 0 0 1 0 0 0 0]);
Df=e1*dfx1+e2*dfx2+e3*dfx3;fD=dfx1*e1+dfx2*e2+dfx3*e3;
DfD=e1*dfx11*e1+e1*dfx12*e2+e1*dfx13*e3+e2*dfx12*e1+e2*dfx22*e2+e2*dfx23*e3
+e3*dfx13*e1+e3*dfx23*e2+e3*dfx33*e3
Lcf=alphaC*DfD+betaC*D2f
Lcphif=alphaC*DphifDphi+betaC*D2f
Lcpsif=alphaC*DpsifDpsi+betaC*D2f
Lgf=alphaC*DphifDpsi+betaC*DphiDpsif
Lgif=alphaC*DpsifDphi+betaC*DpsiDphif
end
\end{verbatim} 
{\bf Example (\eqref{ejemplo}-\eqref{study})}
\begin{verbatim}
>> Lame([-1 0 0],[0 1 0],[0 0 1],[1 0 0],[0 1 0],[0 0 1],[0 x1*x2 
-2*x1^2-3*x2^2+5*x3^2 x3 0 0 0 0],[0.1 0.2])
Coefficients meet Lame`s restrictions
D2f = 0e0   DphiDpsif = 0e0 + -10e2   DpsiDphif = 0e0 + -6e2
DphifDpsi = 0e0 + 20e2   DpsifDphi = 0e0 + 20e2   DphifDphi = 0e0 + 14e2
DpsifDpsi = 0e0 + 10e2   DfD = 0e0 + 10e2
Lcf = 0e0 + 1e2   Lcphif = 0e0 + 7/5e2   Lcpsif = 0e0 + 1e2
Lgf = 0e0   Lgif = 0e0 + 4/5e2
\end{verbatim}
}}

\end{document}